\documentclass[aps,prl,showpacs,twocolumn,superscriptaddress,
amsmath,floatfix]{revtex4-1}
\usepackage{graphicx}
\usepackage{epstopdf}
\usepackage{tikz}
\usepackage{verbatim}
\usepackage{bm}
\usepackage{dcolumn}
\usepackage{color}
\definecolor{gray}{rgb}{0.8,0.8,0.8}
\definecolor{gr}{rgb}{0,0.6,0}
\usepackage[colorlinks,bookmarks=false,citecolor=blue,linkcolor=red,urlcolor=blue]{hyperref}

\usepackage{bm}
\usepackage[all,arc]{xy}
\usepackage{amsmath,amssymb}
\usepackage{cancel}
\usepackage{wasysym}

\usepackage{booktabs}

 \DeclareRobustCommand{\singD}{
  \begin{tikzpicture}
    \filldraw[color=gray,fill=gray,thick](-8.4,-0.09)  rectangle (-8.0,0.09);
    \filldraw[color=black, fill=white](-8,0) circle (0.1);
    \filldraw[color=black, fill=white](-8.4,0) circle (0.1);
     \end{tikzpicture}}

\begin{document}

\title{Spin-Orbit Dimers and Non-Collinear Phases in $d^1$ Cubic Double Perovskites}

\author{Judit Romh\'anyi}
\affiliation{Max Planck Institute for Solid State Research, Heisenbergstrasse 1, D-70569 Stuttgart, Germany}
\affiliation{Okinawa Institute of Science and Technology Graduate University, Onna-son, Okinawa 904-0395, Japan}
\author{Leon Balents}
\affiliation{Kavli Institute for Theoretical Physics, University of California, Santa Barbara, CA 93106, USA}
\author{George Jackeli}
\altaffiliation[]{Also at Andronikashvili Institute of Physics, 0177
Tbilisi, Georgia}
\affiliation{Max Planck Institute for Solid State Research,
Heisenbergstrasse 1, D-70569 Stuttgart, Germany}
\affiliation{Institute for Functional Matter and Quantum Technologies, 
University of Stuttgart, Pfaffenwaldring 57, D-70569 Stuttgart, Germany}
\date{\today}

\begin{abstract}
We formulate and study a
spin-orbital model for a family of cubic
double perovskites with $d^1$ ions occupying a frustrated fcc
sublattice. A variational approach and a complimentary 
analytical analysis reveal a rich variety of phases emerging from
the interplay of Hund's and spin-orbit couplings (SOC). The phase digram 
includes non-collinear ordered states, with or without net  
moment, and, remarkably, 
a large window of a non-magnetic disordered  spin-orbit dimer phase.
The present theory uncovers the physical origin of the unusual amorphous valence bond state experimentally suggested for
Ba$_2${\it B}MoO$_6$ ({\it B}=Y,Lu), and predicts possible ordered patterns in
Ba$_2${\it B}OsO$_6$ ({\it B}=Na,Li) compounds.
\end{abstract}

\maketitle

Conventionally, frustration, low dimensionality and low spin are the key 
attributes of emerging novel quantum ground states. 
In the quest to realize a quantum spin liquid, a state of spins
possessing massive quantum entanglement and lacking magnetic order,
researchers have extensively studied
Mott insulators  with antiferromagnetic (AF) interactions on geometrically frustrated triangular, kagome, hyper-kagome 
and pyrochlore lattices~\cite{Balents2010,Savary2016}.
Another route to frustration in Mott insulators with unquenched angular momentum is provided by
orbital degrees of freedom. The directional character of degenerate $d$-orbitals 
may frustrate the magnetic interactions even on bipartite lattices,  and lead to a 
plethora of emergent phases with unusual spin patterns ~\cite{Kugel1982,Khaliullin2005} or without 
long-range spin/orbital order~\cite{Pen1997,Khaliullin2000,Vernay2004,diMatteo2004,diMatteo2005,Jackeli2008}.

In $4d$ and $5d$ transition metal compounds, the enhanced SOC, compared to $3d$ systems, 
fully or partly lifts the local  degeneracy of a $d$-shell. When degeneracy is fully lifted, e.g. in case of  a single hole 
in a $t_{2g}$-shell, the anisotropic orbital interactions as well as related frustration are transferred to pseudo-spin 
one-half Kramers doublets  of $d^5$ ions~\cite{Khaliullin2005,Chen2008,Jackeli2009}. However, in case of only 
partially lifting the degeneracy, the directional character of the electron density of the degenerate states is preserved, 
resulting in an effective reduction of magnetic sublattice dimensionality and strongly amplifying the effects of geometrical 
frustration. The Mott insulating $d^1$ double perovskites with undistorted cubic structure, such as spin-1/2
Ba$_2${\it B}MoO$_6$ ({\it B}=Y,Lu) and Ba$_2${\it B}OsO$_6$ ({\it B}=Na,Li), in which the only magnetically active ions, 
Mo$^{5+}$ or Os$^{7+}$, reside on a weakly frustrated fcc sublattice well exemplify this physical scenario~\cite{Chen2010}. 

The osmium compounds  Ba$_2$NaOsO$_6$ and Ba$_2$LiOsO$_6$  order 
magnetically~\cite{Stitzer2002,Erickson2007,Steele2011}.  Small effective local moments $\sim\!0.7\;\mu_{\rm B}$, 
compared to spin only value 1.7~$\mu_{\rm B}$, have been extracted from high temperature susceptibilities in both 
materials~\cite{Stitzer2002}. The strong reduction of local moments is a direct manifestation of  unquenched orbital 
momentum and strong SOC in the $5d$-shell of Os$^{7+}$ ion~\cite{Pickett2007,Pickett2015,Pickett2016}.
In Ba$_2$NaOsO$_6$, anomalously small net ordered moment
$\sim\!0.2\mu_{\rm B}$ has additionally been
detected~\cite{Erickson2007,Steele2011}.   Recent NMR measurements
indicate a canted AF order in the Na compound~\cite{lu16:_obser_novel_magnet_local_symmet}.

The reported experimental data on Ba$_2$YMoO$_6$ are even more puzzling:
this compound does not show any structural or magnetic transition down to 
$50$~mK~\cite{Aharen2010,deVries2010,deVries2013}. The total high
temperature entropy extracted from electronic heat capacity was reported to be 
close to $R\ln 4$~\cite{deVries2010}, indicating the presence of an extra 
two-fold orbital degeneracy in addition to the spin, and allowing for the 
emergence of multi-orbital physics. Based on magnetic susceptibility and muon spin
rotation data, a valence bond glass state, an amorphous
arrangement of spin singlets, has been proposed for  
Ba$_2$YMoO$_6$~\cite{deVries2010} which remains quite stable against
 isovalent substitutions of Ba$^{2+}$ with Sr$^{2+}$~\cite{Mclaughlin2010}.  
The magnetic susceptibility of a very similar compound Ba$_2$LuMoO$_6$ also did not
exhibit  any magnetic transition  down to 2~K~\cite{Coomer2013}.  
Theoretically, various exotic phases, including multipolar order~\cite{Chen2010} and 
chiral spin-orbital liquid~\cite{Natori2016}, have been put forward as possible candidates.

In this letter, we introduce and study a spin-orbital model and show that a  dimer-singlet phase, 
composed of random arrangement of spin-orbit dimers,
without any type of long-range order is a natural ground state of the model. The physical properties of 
this disordered phase  are consistent with all available experimental findings on molybdenum  double perovskites.
In addition, the minimal model supports complex non-collinear, coplanar, ordered patterns. We argue that
such four-sublattice ordered states are realised in osmium compounds.

\begin{figure}[!htp]
\begin{center}
\includegraphics[width=0.85\columnwidth]{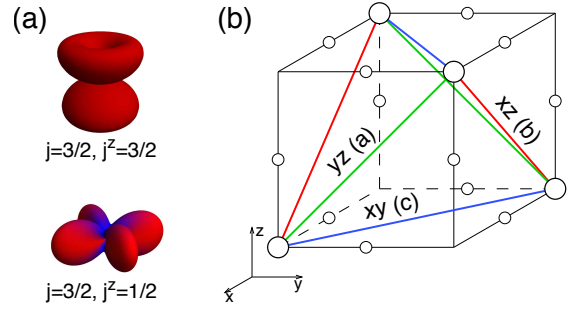}
\caption{(a) Density profile of  $j=3/2$ quartet. The states $j^z=\pm\frac{1}{2}$ (bottom) have dominant $xy$-orbital character, 
while $j^z=\pm\frac{3}{2}$ components (top) are composed of $xz$ and $yz$ orbitals. Red and blue coloring denotes the up 
and down spin distribution, respectively. (b) Crystallographic unit cell containing four molybdenum ions (large circles). Oxygen 
positions are indicated by small circles. The  nearest-neighbor bonds  belonging to different cubic planes are distinguished by 
different colors. The $t_{2g}$-orbitals active along the corresponding bonds are also indicated.}
\label{fig:fig1}
\end{center}
\end{figure}

{\it Local electronic structure}.-- The single $d$-electron of a Mo$^{5+}$ or Os$^{7+}$ ion in a cubic environment 
occupies $t_{2g}\text{-manifold}$ of degenerate $xy$, $xz$, $yz$ orbitals. It carries an effective angular momentum $l=1$ with 
$|l^z\!\!=\!\!0\rangle \equiv \!\!|xy\rangle$, $|l^z\!\!=\!\!\pm 1\rangle \equiv \!\!-\frac{1}{\sqrt{2}}(i|xz\rangle\pm|yz\rangle)$~\cite{Abr70}.  
The six-fold degeneracy of the local Hilbert space is lifted by the local SOC $H_{\rm so}=-\lambda {\vec l}\cdot{\vec S}$ 
stabilizing $j=l+S=\frac{3}{2}$ quartet  and pushing $j=\frac{1}{2}$ Kramers doublet to
a higher energy. Here, ${\vec S}$ is an electron spin operator and $\lambda$ denotes the SOC. 
The  states $j^z=\pm\frac{1}{2}$  of $j=\frac{3}{2}$ manifold  have predominantly $xy$ character, while $j^z=\pm\frac{3}{2}$ 
components are given by superposition of $xz$ and $yz$ orbitals only [see Fig.~\ref{fig:fig1}(a)]. When SOC
is much smaller  (larger) than the exchange interactions between 
neigbhoring  ions, it is more convenient to use the $t_{2g}$ ($j=\frac{3}{2}$) basis. 
The following analysis covers both limits. 

{\it Spin-orbital Hamiltonian}.-- In the double perovskite structure, each nearest-neighbor bond of the 
fcc sublattice of magnetic ions belongs to one of the crystallographic planes $xy$, $xz$, or $yz$ as shown in  
Fig.~\ref{fig:fig1}(b). We label these bonds as well as the $t_{2g}$-orbitals with a cubic axis $\gamma(=a,b,c)$ 
normal to their planes, e.g. $xy$ becomes $c$. The hopping between neighboring $t_{2g}$-orbitals takes place through
intermediate oxygens' $p$-orbitals, or direct hybridization. Along
a $\gamma$-type bond the dominant overlap, with amplitude $t$, is between
$\gamma$-orbitals~\cite{Chen2010, Streltsov}. The low-energy spin-orbital model is obtained via standard second order perturbation
theory in $t/U$ ($U$ being the local Coulomb repulsion)~\cite{tU}, and reads as follows:
\begin{eqnarray}
{\cal H}&=&\sum_{\langle ij\rangle_{\gamma}}{\Bigr [} -J_1{\bigr (}\vec S_i\cdot \vec S_{j}+\frac{3}{4} {\bigl )}
+J_2{\bigr (}\vec S_i\cdot \vec S_{j} -\frac{1}{4} {\bigl
  ){\Bigl ]}}P_{ij}^{(\gamma)}
\nonumber\\
&+&J_3\sum_{\langle ij\rangle_\gamma} {\bigr (}\vec S_i\cdot \vec S_{j}-\frac{1}{4} {\bigl
)}\bar{P}_{ij}^{(\gamma)}
-\lambda\sum_i{\vec l}\cdot{\vec S}~.
\label{eq:eq1}
\end{eqnarray}
 ${\langle ij\rangle_{\gamma}}$ denotes a $\gamma$-type  bond, $J_{1(2)}=\frac{1}{4}Jr_{1(2)}$, $J_3=\frac{1}{3}J(2r_2+r_3)$,
$J=4t^2/U$, the set of $r_n$ describing  the multiplet 
structure of excited states are functions  of  $\eta=J_{H}/U\ll 1$~\cite{rvals}, and $J_{H}$ is  the  Hund's coupling. 

The isotropic spin exchange couplings  depend on the orbital occupancy of the corresponding bonds~\cite{Kugel1982,SM}, 
and are described by the first three terms of Eq.~(\ref{eq:eq1}), with the orbital projectors 
$P_{ij}^{(\gamma)}=n_{i}^{(\gamma)}(1-n_{j}^{(\gamma)})+(1-n_{i}^{(\gamma)})n_{j}^{(\gamma)}$ and
$\bar{P}_{ij}^{(\gamma)}=n_{i}^{(\gamma)}n_{j}^{(\gamma)}$, where 
$n_{i}^{(\gamma)}$ is the occupation number of a $\gamma$-orbital.
The spin isotropy is broken by the SOC in Eq.~(\ref{eq:eq1}),
allowing symmetric  anisotropic exchange
between  $j=\frac{3}{2}$ quartets.  In cubic double perovskites, the antisymmetric Dzyaloshinsky-Moriya 
exchange is forbidden by  the bond inversion symmetry.

{\it Dimer-singlet phase}.-- We start our analysis by 
setting the small parameter  $\eta=0$, and discuss later the model~(\ref{eq:eq1}) in its full parameter  space.
We consider  two limiting cases when $\lambda\ll J$ or $\lambda\gg J$, and identify the ground state phases of the  model~(\ref{eq:eq1}) through analytical considerations.  At $\eta=0$, first three terms of the model~(\ref{eq:eq1}) can be grouped, up to a constant term, 
into one ~\cite{SM}, and the model simplifies to
 \begin{eqnarray}
 \mathcal{H}=J\sum_{{\langle ij \rangle}_{\gamma}}{\bigr (}\vec S_i\cdot \vec S_{j}+\frac{1}{4} {\bigl )}\bar{P}_{ij}^{(\gamma)}-\lambda\sum_i{\vec l}\cdot{\vec S}~.
\label{eq:eq2}
\end{eqnarray}
The expectation value of the first term in Eq.~(\ref{eq:eq2}) in any classical, i.e. site-factorized, state is
non-negative. At $\lambda=0$, the zero minimum classical energy is achieved by forming decoupled layers of AF square lattices with uniform planar orbital order. In this state, the orbital projectors $\bar{P}_{ij}^{(\gamma)}=1(0)$ 
on intra-(inter-)layer bonds and $\langle {\vec S}_i\cdot{\vec S}_j\rangle=-\frac{1}{4}$ on intra-layer bonds. Hence, orbital `flavors' are decoupled and flipping locally an orbital 'flavor' does not cost  
energy, resulting in a massive  ground state degeneracy~\cite{SM}. A product state constructed from entangled quantum 
spin-orbit states on decoupled dimer bonds has however lower negative energy, ${\sf E}_{\sf DS}=-\frac{1}{4}J$. This phase, 
termed here as dimer-singlet phase,  corresponds to a hard-core dimer covering of the fcc lattice, with  $\bar{P}_{ij}^{(\gamma)}=1(0)$  on  (inter-)dimer bonds. On a dimer bond, 
spins form a singlet and occupied orbitals have lobes directed along the bond. Covering the lattice with such dimers 
 is in fact an exact eigenstate of the Hamiltonian~(\ref{eq:eq2}).  When neighboring dimers are in the same plane, an energetically 
unfavorable  larger clusters of AF coupled spins are formed~\cite{Jackeli2007a}, and such configurations are banned 
from the ground state manifold. Although, this seems to be a rather strong constraint, the orientational 
degeneracy of dimer covering remains extensive~\cite{SM}.

For  $\lambda\gg J$, the $t_{2g}$-levels are split and the components of the lower $j=3/2$ quartet forms the relevant 
basis, that we label by pseudo-spin ${\vec s}$ and pseudo-orbital $\vec{\tau}$ states: 
$|\tau^z\!=\!\frac 1 2\!,~\!s^z\!=\!\pm\frac 1 2\rangle\!\equiv\!|j\!=\!\frac 3 2,~j^z\!=\!\pm\frac 1 2\rangle$ and 
$|\tau^z\!=\!-\frac 1 2\!,~\!s^z\!=\!\pm\frac 1 2\rangle\!\equiv\!|j\!=\!\frac 3 2,~j^z\!=\!\pm\frac 3 2\rangle$~\cite{natori}.
Projecting Eq.~(\ref{eq:eq2}) onto this new basis, we find 
\begin{eqnarray}
\mathcal{H}={\tilde J}\sum_{{\langle ij \rangle}_{\gamma}} {\bigr (}\vec s_i\cdot \vec s_{j}+\frac{1}{4} 
{\bigl )}\tilde{P}_{ij}^{(\gamma)}~,
\label{eq:eq3}
\end{eqnarray}
where $\tilde{P}_{ij}^{(\gamma)}=(\frac 1 2+ \tau^{(\gamma)}_i )(\frac 1 2+ \tau^{(\gamma)}_j)$, ${\tilde J}=\frac{4}{9}J$, $\tau^{(a)}=-\frac{1}{2}\tau^z-\frac{\sqrt{3}}{2}\tau^x$,  $\tau^{(b)}=-\frac{1}{2}\tau^z+\frac{\sqrt{3}}{2}\tau^x$, 
and $\tau^{(c)}=\tau^z$.  Hamiltonian~(\ref{eq:eq3}) has the same form as the Kugel-Khomskii model of $e_g$-orbitals 
on a cubic lattice~\cite{Kugel1982} and explicitly reveals the emergent, at large $\lambda$,   hidden SU(2) symmetry pointed out  
in Ref.~\onlinecite{Chen2010}. Similarly to  $\lambda=0$,  the ground state manifold of~(\ref{eq:eq3}) is spanned by dimer-singlets, but now these are composed  of pseudo-spins instead of real spins.

Insight for intermediate $\lambda$ can be gained by  exactly solving the model~(\ref{eq:eq1}) on an isolated bond, since the inter-dimer couplings appear to be much smaller than intra-dimer ones (see below). For each values of $\lambda$, we find the singlet ground state
\begin{eqnarray}
\left|\singD\right>=\bigr(\left|\Uparrow\Downarrow\right>-\left|\Downarrow\Uparrow\right>\bigl)/\sqrt{2}\;,
\label{eq:singlet}
\end{eqnarray}
where the wave-functions of pseudo-spins $\Uparrow(\Downarrow)$  depend on the
strength of $\lambda$~\cite{SM}, e.g., in the $xy$-plane,  we have
\begin{eqnarray}
\left|\Uparrow(\Downarrow)\right>=\cos\vartheta\left|0,\uparrow(\downarrow)\right>+\sin\vartheta\left|(-)1,\downarrow(\uparrow)\right>~.
\label{eq:pseudo}
\end{eqnarray}
In the two limiting cases, $\lambda=0$ and $\lambda\gg 1$, the variational parameter $\theta$ becomes $0$ and 
$\arccos\sqrt{\frac{2}{3}}$, respectively. The SOC inflates the planar orbital, so that at large $\lambda$ it becomes  $\left|j\!=\!\!\frac{3}{2},j_z\!\!=\!\!\pm\!\frac{1}{2}\right>$. The latter has small out-of-plane component, see Fig.~\ref{fig:fig1}(a), generating finite  but small interactions between, otherwise decoupled,  dimers.  However, as it follows, 
inter-dimer couplings do not select any particular superstructure of dimers.
\begin{figure}
\begin{center}
\includegraphics[width=1.0\columnwidth]{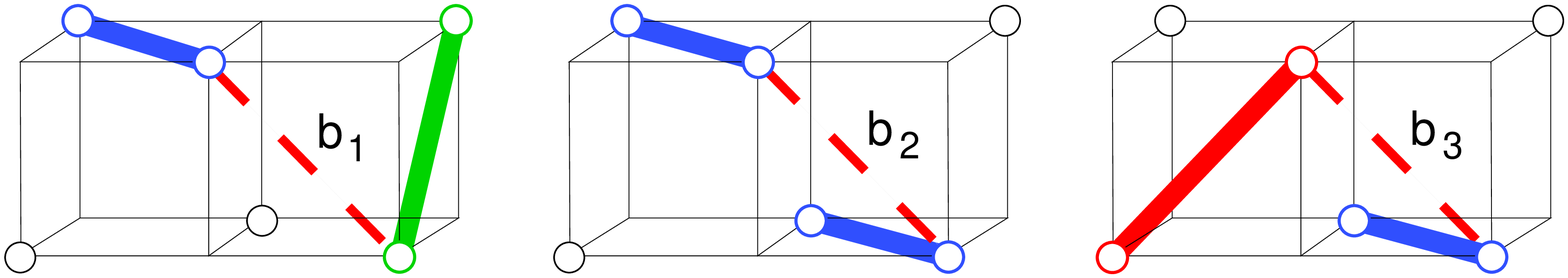}
\caption{Types of different inter-dimer bonds depicted as dashed lines. Dimers are represented with $\singD$ and are colored according  to three cubic planes they belong to.
Bonds ${\sf b_1}$ and ${\sf b_2}$ couple dimers with perpendicular to these bonds. Bond ${\sf b_3}$ connects a dimer  
from a orthogonal plane with   another one that is in the same plane as ${\sf b_3}$ itself.}
\label{fig:fig2}
\end{center}
\end{figure}

Fig.~\ref{fig:fig2} shows  all possible inter-dimer bonds allowed in the ground state manifold.  Such a bond may connect two dimers 
both perpendicular
 the connecting bond itself: then, either the connected dimers belong to different planes $({\sf b_1})$ or to the same plane $({\sf b_2})$. 
The third possibility, ${\sf b_3}$,   is that one of the dimers is in the same plane as the inter-dimer bond,
and the other is perpendicular to them [see Fig.~\ref{fig:fig2}]. Consequently, regardless of the dimer arrangements, each dimer has exactly six neighboring ${\sf b_3}$ bonds.  
Out of 6${\cal N}$ bonds of the fcc lattice with ${\cal N}$ sites, there are  $\frac{1}{2}{\cal N}$ 
dimer and  $3{\cal N}$  of ${\sf b_3}$-type bonds, thus remaining $\frac{5}{2}{\cal N}$ bonds are  ${\sf b_1}$-  or ${\sf b_2}$-types. 
Each dimer (${\sf b_n}$-type) bonds host a finite energy ${\cal E}_{\sf d}$ (${\cal E}_{\sf b_n}$).  As  both ${\sf b_1}$ and ${\sf b_2}$ bonds connect dimers out of their plane, ${\cal E}_{\sf b_1}={\cal E}_{\sf b_2}$ and the energy of a product dimer state 
\begin{equation}
{\sf E}_{\sf DS}=\bigl({\cal E}_{\sf d}+5{\cal
  E}_{\sf b_1}+6{\cal E}_{\sf b_3}\bigr){\cal N}/2
\label{eq:grse_dimer}
\end{equation}
is independent of the dimer covering. Hence, the inter-dimer couplings do not order dimers  
and the massive orientational degeneracy persists. In real materials, however, a mis-site disorder 
and/or uncorrelated local distortions  most likely select a random dimer covering, rendering the system to freeze in a glassy manner. 

\begin{figure}
\begin{center}
\includegraphics[width=0.95\columnwidth]{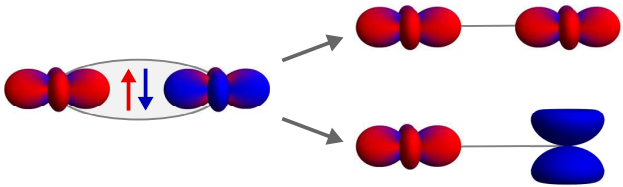}
\caption{Excitations of a spin-orbit dimer in the large $\lambda$ limit. Flipping locally, for example, the $j^z=-1/2$  component 
to $j^z=1/2$ or to $j^z=-3/2$ corresponds to flipping the pseudo-spin (top) or pseudo-orbital (bottom), respectively. Local excitations 
for small SOC correspond to changing the real spins or orbitals.}
\label{fig:fig3}
\end{center}
\end{figure}

In an amorphous dimer-singlet phase, momenta of the excitations are not well defined, but their energies are.  Moreover,
the inter-dimer couplings are much smaller than  the intra-dimer exchange,
allowing  isolated dimer description of the bulk magnetic spectra.
At $\eta,\lambda=0$, as  product dimer states are exact eigenstates,  spins of different dimers are completely decoupled. In the large $\lambda$ limit, the inter-dimer pseudo-spin exchange  ${J}^\prime\simeq \frac{1}{16} {\tilde J}\ll {\tilde J}$. This estimate  follows from Eq.~(\ref{eq:eq3}) by noting that $\langle\tilde{P}_{ij}^{(\gamma)}\rangle=\frac{1}{16}$ on the inter-dimer bonds.
Two types of local excitations allowed by magnetic dipole transitions are illustrated in Fig.~\ref{fig:fig3}.  The upper one corresponds to flipping 
locally a (pseudo-)spin at the energy cost $\Delta_\text{S}= J\;({\tilde J})$ in small (large) $\lambda $ limit. 
The lower is a (pseudo-)orbital excitation  that costs half 
the energy, $\Delta_\text{O}= \frac{1}{2}J\;( \frac{1}{2}{\tilde J})$, of a spin-like excitation. These estimates follow  from the expectation values of the limiting Hamiltonians~Eqs.(\ref{eq:eq2},\ref{eq:eq3}) in the ground state  of an isolated bond, Fig.~\ref{fig:fig3}(left), and  its excited states, Fig.~\ref{fig:fig3}(right). Using reported parameters for  Ba$_2$YMoO$_6$~\cite{tU}, we estimate energy of spin-like (orbital-like) excitations $\Delta_\text{S(O)}\simeq 20-45\;(10-23)$~meV, for large$-$small SOC, and their bandwidth
($\sim J^\prime$) of about few meV.  In the magnetic dipolar channel,  spin-like excitations carry stronger intensity 
than orbital-like ones.  These  findings agree well with neutron scattering data on powder samples  discussed below. 

There are additional thermally accessible non-local excitations at lower energies. For example AF coupled 
spin clusters, or orphan spins  may emerge as a result of thermally induced orbital reorientation.  An important difference between the 
well studied spin-only dimer systems and our model is the lack of a hard-gap. Here, on account of orbital degrees of freedom,
the spectrum cannot be characterized by a  single energy scale.

\begin{figure}
\begin{center}
\includegraphics[width=0.95\columnwidth]{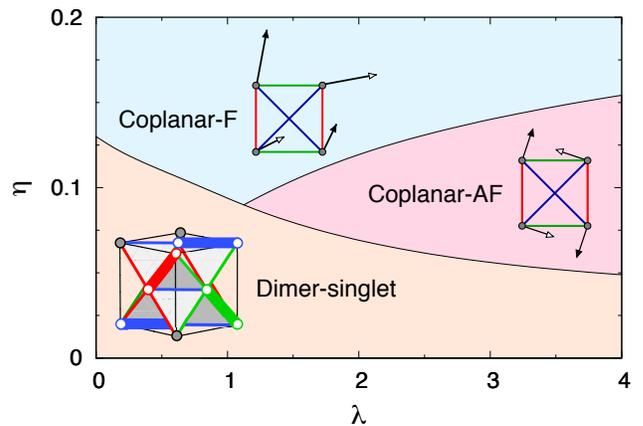}
\caption{Phase diagram of the model~(\ref{eq:eq1}) as the function of Hund's coupling $\eta$ and the SOC $\lambda$ (in units of $J$). 
For small values of  $\eta$ the dimer-singlet phase (see inset) is stable over the entire range of $\lambda$. With increasing  Hund's coupling, 
non-collinear coplanar phases with ordered moments in one of the cubic planes are stabilized.  The ferro-type coplanar state, coplanar-F,  
has  a finite net moment pointing along $[110]$ (or equivalent) direction. The cartoon figures show a tetrahedron of four molybdenum sites 
projected onto the plane of ordered  ${\vec j}$-moments, depicted as arrows.}
\label{fig:fig4}
\end{center}
\end{figure}

{\it Phase diagram}.-- 
To explore the entire phase diagram of the full Hamiltonian~(\ref{eq:eq1}),  we used a site-factorized variational approach and compared
the energies of ordered and dimer-singlet phases. The latter is numerically obtained from  Eq.~(\ref{eq:grse_dimer}) using a product state of the exact  wave-functions of isolated dimers. 
Within  our variational approach, the magnetic and crystallographic unit cells coincide, however, we still need forty variational 
parameters to construct a trial wave-function~\cite{SM}.  When $\eta=0$ the ground state
is a random arrangement of spin-orbit dimers [see inset in Fig.~\ref{fig:fig4}] for any value of $\lambda$. Only the nature of pseudo-spins forming the singlet
dimers is affected by $\lambda$, in accordance with the above analytical considerations.  
For large enough Hund's coupling, we find two non-collinear but coplanar
phases of ordered total angular momenta ${\vec j}$ [see Fig.~\ref{fig:fig4}].  One, termed here as  coplanar-F, has finite net moment along  $[110]$  (or equivalent) direction, i.e. along one of the NN bond, as  experimentally observed~\cite{Stitzer2002}.  The other,  coplanar-AF, has no net moment.
In the dimer-singlet phase, on a dimer bond in $\gamma$-plane corresponding $\gamma$-orbital is predominately occupied, with  occupancy
decreasing from $n^{(\gamma)}=1$ to $\frac{2}{3}$ with increasing $\lambda$. Hunds coupling induced transitions to ordered states are accompanied by  complex rearrangements of an electron density within  SOC split $t_{2g}$-multiplet, with orbital occupancies  dictated by the actual values of  parameters, e.g.  in coplanar-AF order in  a cubic
$\gamma$-plane  the   $\alpha$- and $\beta$-orbitals are predominantly occupied compared to the in-plane $\gamma$-orbital. All phase boundaries appear to be first order within our approach: the net moment  and the order parameters drop to zero across the transitions from coplanar-F to coplanar-AF state and  from the ordered  to disordered dimer-singlet phase, respectively. However, one cannot rule out a second order symmetry allowed transition between ordered states,  or an exotic continuous transition from spontaneously  dimerized phase to ordered states~\cite{Senthil2004}.

{\it Experimental implications}.-- The dimer-singlet phase captures  experimental observations on
the molybdenum compounds. In agreement with experiments, it does not exhibit any  long-range ordering nor breaks any global symmetry.  Its 
extensive degeneracy explains the observed  glassy behavior and suggest the presence of  a residual entropy, that cannot be excluded based on heat capacity data~\cite{deVries2010}. 
Magnetic susceptibility and electronic heat capacity~\cite{deVries2010,deVries2013} 
suggest the  presence  of pseudo-gapped, rather than hard-gapped, low-energy excitations, consistent with the dimer-singlet phase.
Neutron scattering experiments on powder samples~\cite{Carlo2011} revealed excitations that are in
line with the spectrum of weakly coupled spin-orbit dimers. An intense  'mode' observed at $\Delta_{\text S}\simeq 28$~meV with bandwidth of about $4$~meV  is interpreted here as a (pseudo-)spin singlet-to-triplet excitation.  A less intense, lower-energy 
($\Delta_{\text O}\simeq9-17$~meV) response  centred around at half the energy of $\Delta_{\text S}$ is naturally attributed  to (pseudo-)orbital excitation. 
These lower-lying excitations have also  been observed  in NMR response~\cite{Aharen2010}.
The energetics of the observed excitations agrees well with above estimates $\Delta_\text{S(O)}\simeq20-45\;(10-23)$~meV.
In addition, the infrared transmission spectra indicate the emergence of uncorrelated  local distortions of MoO$_6$ octahedra
below 130~K~\cite{Qu2013}, at around the same temperature the magnetic susceptibility start to decrease, most likely due to 
formation  of spin-orbit dimers. 
In the dimer-singlet phase, such uncorrelated distortions  emerge due to  the directional character of the occupied orbitals.

The four-sublattice ordered states in the phase diagram (Fig.~\ref{fig:fig4})
may provide description for the iso-structural osmium compounds, Ba$_2$LiOsO$_6$ and Ba$_2$NaOsO$_6$.
The latter is characterized  by very small net magnetic moment $\sim\!0.2\mu_{\rm B}$ along $[110]$ easy axis~\cite{Stitzer2002}.
We find  the net  moment
$\vec{M}=2\vec{S}-\vec{l}$ along the same  $[110]$ (or equivalent) direction, being $\sim 1 \mu_B$ for small and $\sim 0.1\mu_B$ for
large $\lambda$.  

To summarize, within  a minimal microscopic model, we have proposed unified theoretical description of possible ground states in $d^1$ cubic double perovskites.
The obtained spin-orbital model shows a rich phase behavior
including a massively degenerate dimer-singlet manifold, without any long-range order,  and unusual non-collinear
ordered patterns. Our theoretical study elucidates physics behind and provides explanations of experimental data on 
molybdenum and osmium based compounds. The physics discussed here
 may also be relevant to other heavy transition metal  compounds,
such as molybdenum pyrochlores,  in which random distribution of `dimerized' bonds, induced by orbital degrees, have been recently revealed by pair-distribution function 
measurements~\cite{Thygesen2016}.

We thank G. Chen, M. Haverkort,  G. Khaliullin, F. Mila, W. Natori,  J.A.M. Paddison, S. Streltsov,  and H. Takagi for discussions. J.R. acknowledges funding from the Hungarian OTKA grant K106047. Work by L.B. was supported by the DOE, Office of Science, Basic Energy Sciences under award number DE-FG02-08ER46524. 
G.J. benefitted from the facilities of the KITP, and was supported in part by the NSF under Grant No. NSF PHY11-25915.

\bibliographystyle{apsrev4-1}
\bibliography{DP_MT_v2}

\clearpage

\pagebreak
\widetext
\begin{center}
\textbf{\large Supplemental Material:\\
``Spin-Orbit Dimers and Non-Collinear Phases in $d^1$ Cubic Double Perovskites''}
\end{center}
\setcounter{equation}{0}
\setcounter{figure}{0}
\setcounter{table}{0}
\setcounter{page}{1}
\makeatletter    
     
\renewcommand{\theequation}{S\arabic{equation}}
\renewcommand{\thefigure}{S\arabic{figure}}
\renewcommand{\bibnumfmt}[1]{[S#1]}
\renewcommand{\citenumfont}[1]{S#1}

\subsection{Spin-orbital superexchange}
\begin{figure}[h!]
\begin{center}
\includegraphics[width=0.5\columnwidth]{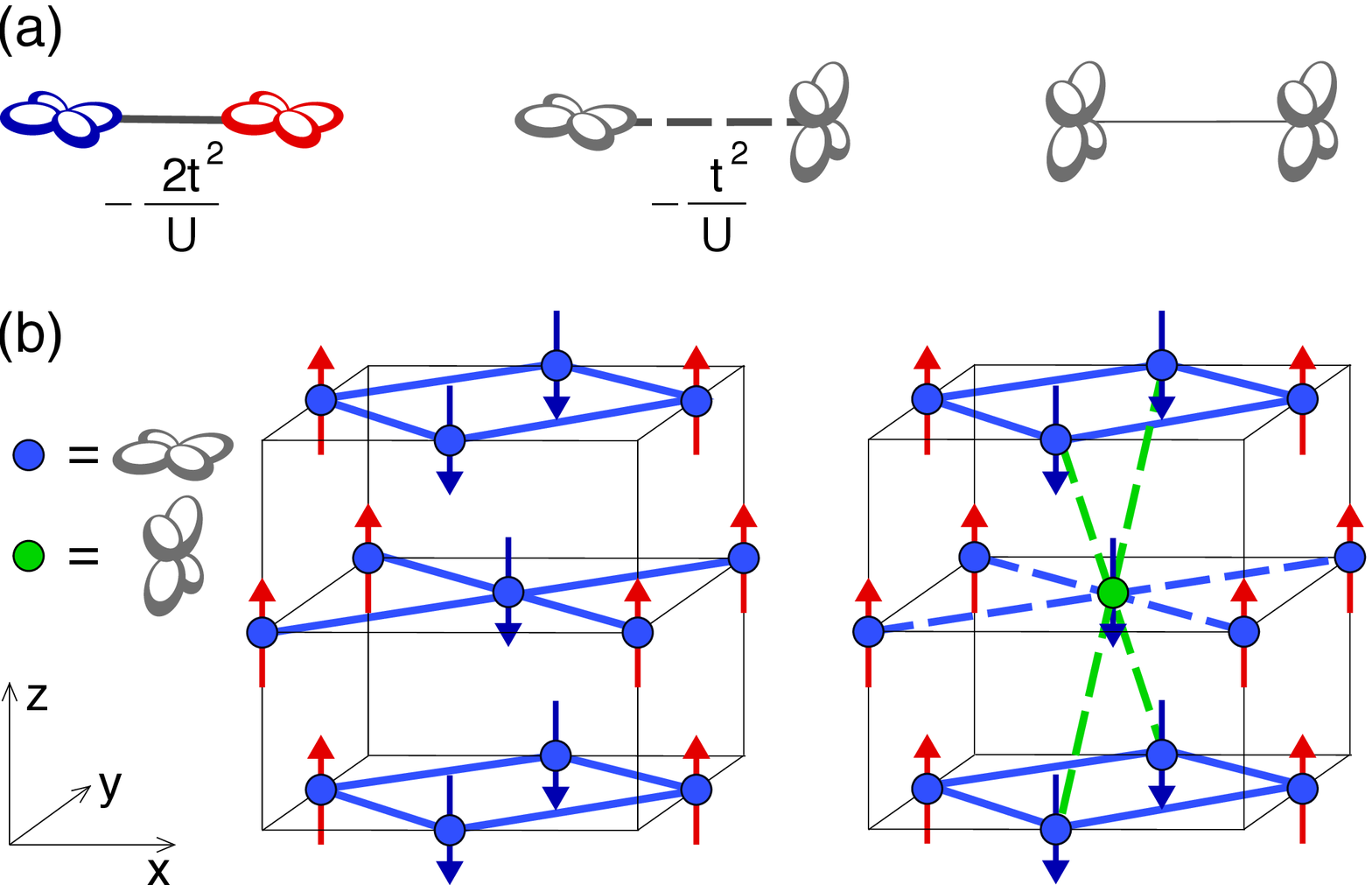}
 \caption{(a) Nearest-neighbor bonds of  $\gamma$-type have zero energy unless one of the electrons occupies  $\gamma$-orbital. The energy gain is maximal when both orbitals correspond to the plane of the bond. (b) The classical ground state is orbitally ordered antiferromagnetic state of decoupled layers (left). Flipping locally an orbital `flavor' does not cost any energy. Although, the bond energies in $xy$-plane are halved, there are new exchange paths to the lower and upper layers (right).
 }
\label{fig:hop_path}
\end{center}
\end{figure}

The first three terms of the  model~(1) in main text describe spin exchange couplings depending on the orbital occupancy and exemplifies  a version of Goodenough-Kanamori rules. 
Three possible orbital configurations of  a bond are shown in  Fig.~\ref{fig:hop_path}(a).
The strongest, antiferromagnetic (AF), exchange  is achieved when the lobs of  occupied orbitals on both neighboring sites point along the bond in between (see Fig.~\ref{fig:hop_path}(a)[left]), and  is given by $J_3$ term in Eq.~(1). When only one occupied orbital is directed along the bond,  Fig.~\ref{fig:hop_path}(a)[middle], the spin exchange 
could be either ferromagnetic (FM) or AF,
$J_1$ and $J_2$ terms in Eq.~(1), respectively. The Hunds coupling induced splitting of virtual doubly occupied state favors the high spin,
$S=1$, configuration and resulting FM exchange is stronger than AF one, $J_1>J_2$. In the limit of zero Hund's coupling, $J_1=J_2$ and spin exchange on such a bond vanishes. When both sites are occupied by orbitals directed away from the common bond, Fig.~\ref{fig:hop_path}(a)[right], there is no virtual hopping process along that bond, hence no  exchange and  energy gain.

\subsection{The limit of zero Hunds coupling}
In the limit $\eta=0$, we are left with the only one exchange scale $J=4t^2/U$, and exchange couplings become  $J_3=4J_2=4J_1=J$. 
The exchange part of  Eq.~(1) can  then be simplified,  by using identities
\begin{eqnarray}
&&\sum_{\gamma}n_{i}^{\gamma}=1~,\;\sum_{{\langle ij \rangle}_{\gamma}}{\bigr (
{P}_{ij}^{(\gamma)}+\bar{P}_{ij}^{(\gamma)}\bigl )}=-\sum_{{\langle ij \rangle}_{\gamma}}\bar{P}_{ij}^{(\gamma)}+4\cal{N}~,\nonumber\\
&&\text{to the form}\;\;\mathcal{H}=J\sum_{{\langle ij \rangle}_{\gamma}}{\bigr (}\vec S_i\cdot \vec S_{j}+\frac{1}{4} {\bigl )}\bar{P}_{ij}^{(\gamma)}~.
\label{eq:Hgamma}
\end{eqnarray}
The classical, site-factorized, ground state manifold of the above  Hamiltonian has an extensive degeneracy due to orbital
degrees of freedom. One of the classical states of the model~(\ref{eq:Hgamma}) on the fcc lattice is built of decoupled antiferromagnetic layers as shown in Fig.~\ref{fig:hop_path}(b).  On the nearest-neighbor bonds of each layer,  the expectation value $\langle {\vec S}_i\cdot{\vec S}_j\rangle=-\frac{1}{4}$  setting the spin part in Eq.~(\ref{eq:Hgamma}) to zero. The ground state energy is then a constant, independent of orbital configurations. Hence, one can locally flip an orbital flavour without changing energy [see  Fig.~\ref{fig:hop_path}(b)] and generate  macroscopically degenerate classical ground state  manifold. 

\subsection{Exact solution of a two-site problem}

\begin{figure}[h!]
\begin{center}
\includegraphics[width=0.4\columnwidth]{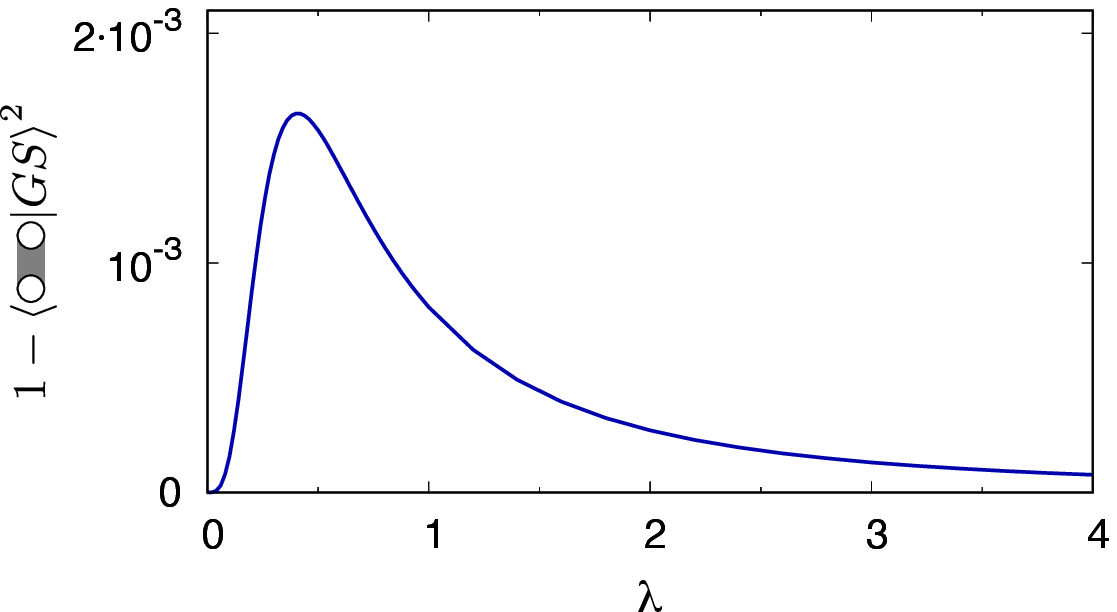}
\caption{Overlap between the exact dimer ground state and the singlet $\left|\singD\right>$~(4) introduced in the main text. 
For arbitrary value of $\lambda$, $\left|\singD\right>$ remains extremely close to the exact solution.
}
\label{fig:overlap}
\end{center}
\end{figure}

To investigate the dimer-factorized solution we begin with the case of an isolated bond, e.g. in $xy$-plane ($c$-type bond). 
The exact ground state wave-function, from numerical diagonalization of the model~(1) of main text, at any values of the spin-orbit coupling $\lambda$,  can be written as linear a combination of two kinds of spin-orbit singlets.
The dominant singlet in this linear combination is   $\left|\singD\right>$ of Eq.~(4) 
with pseudo-spins introduced in Eq.~(5) of the main text.
Contribution from the other singlet vanishes in the limiting cases $\lambda=0$ and $\lambda\to\infty$ and is negligibly small for any value of $\lambda$. Thus, the ground state is approximately a pure singlet  of Eq.~(4). 
To see how close the trial singlet state~(4) is to the exact dimer ground state we plot the overlap between them in Fig.~\ref{fig:overlap}.

\subsection{Extensive degeneracy of constrained dimer coverings}

\begin{figure}[!ht]
\begin{center}
\includegraphics[width=0.5\columnwidth]{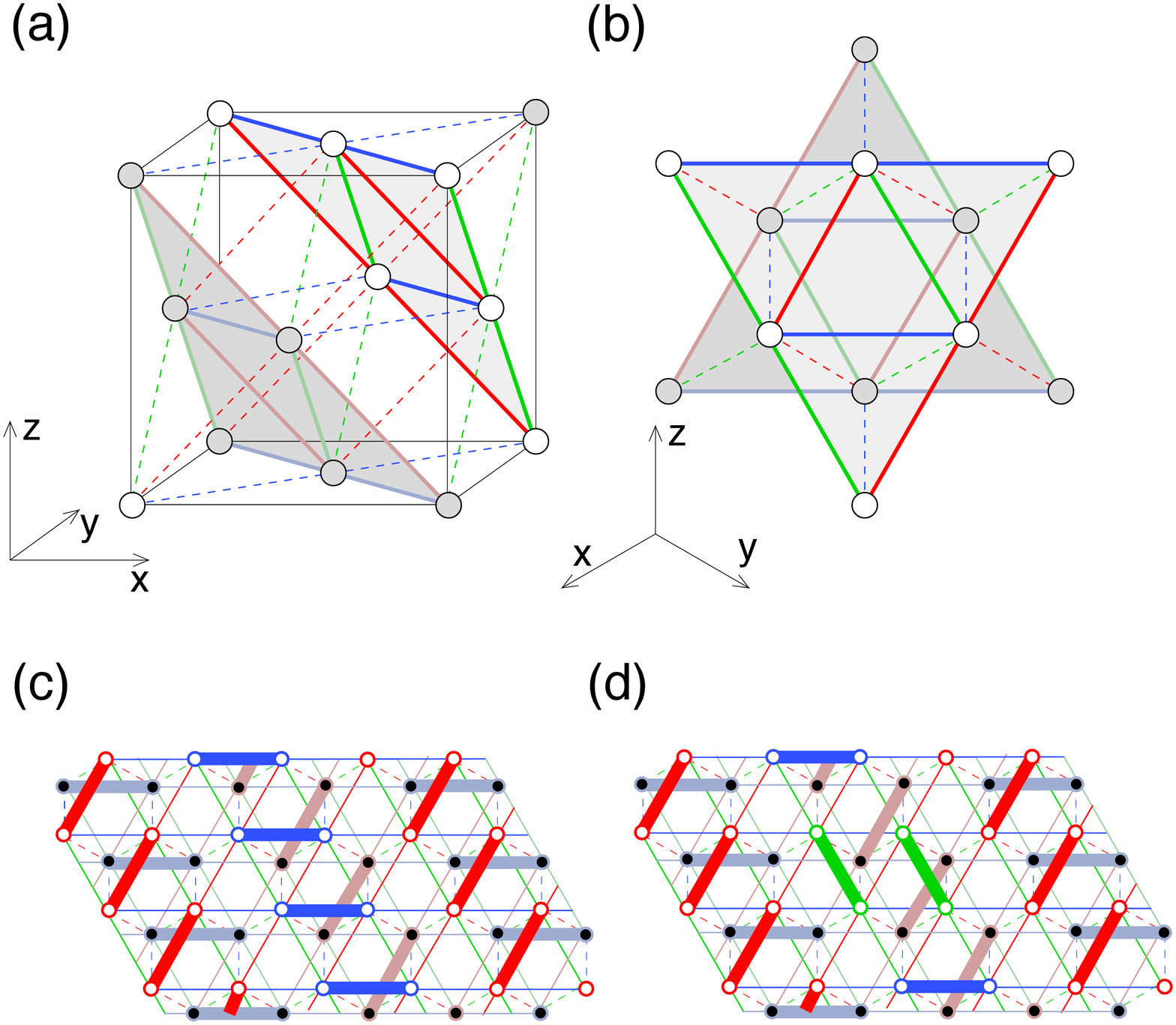}
\caption{(color online)
(a) and (b) illustrate two neighboring $[111]$-planes of the fcc lattice indicated with black and white sites.
(c) A possible  covering of fcc lattice  with  $b$-type (red) and $c$-type (blue) `orthogonal' dimers. The two layers depicted here are denoted with grey and white sites. The solid and dashed lines correspond to intra and inter layer bonds, respectively. (d) Switching the dimers on four neighboring sites in a given layer does not violate the `no-plane' constraint and does not effect the ground state energy.
}
\label{fig:dimer_covering}
\end{center}
\end{figure}

In the following, we show that in spite of the constraint that do not allow two neighbouring dimers to be in the same plane (`no-plane' constraint), there is infinite number of dimer configurations of which the system chooses upon freezing into a glassy disordered phase.
Let us consider the $[111]$-planes of the fcc lattice which naturally contains all three types of bonds ($a$, $b$ and $c$) as shown in Fig.~\ref{fig:dimer_covering}(a). The fcc lattice can be viewed as stacked layers of triangular lattice, where these layers are the $[111]$ planes and the three sides of the triangles correspond to the three kinds of bonds [see Fig.~\ref{fig:dimer_covering}(b)]. A possible dimer-product state is to cover a layer with stripes of two kinds of dimers and repeat the pattern in the neighboring layers so that the different dimers alternate on top of each other.  Two consecutive layers of such a dimer configuration is illustrated in Fig.~\ref{fig:dimer_covering}(c). For convenience we denoted the top layer with bright, and the bottom layer with dim colors. It is easy to see that changing the dimer state of four neighboring sites within a layer does not violate the `no-plane' constrain and hence, does not change the ground state energy. Such a flip is illustrated in Fig.~\ref{fig:dimer_covering}(d). As we can flip anywhere, even successively at more than one plaquette of four sites, there are an extensive number of possible dimer coverings.\\


\subsection{Variational approach}

To compare the dimer-singlet solution with other possible phases we performed site-factorized variational calculations. We look for phases which have a unit cell equivalent to the crystallographic one, and as such, our variational approach cannot describe incommensurate orderings or patterns with larger unit cell. 
As pointed out in the main text, the local Hilbert space of a molybdenum site consists of the six states of a t$_{2g}^1$-configuration. Namely, $|\pm1,\sigma\rangle$ and $|0,\sigma\rangle$, with spin variables $\sigma=\uparrow$ or $\downarrow$.
For simplicity, let us denote these states in the following way; $|1,\uparrow\rangle,\cdots,|-1,\downarrow\rangle=|\phi\rangle_1,\cdots,|\phi\rangle_6$.

In order to find the most general site-factorized solution we 
chose our local variational wave function to be 
\begin{equation}
|\psi_i\rangle=\sum_{n=1}^{6} \zeta_{n} |\phi_i\rangle_n\;,
\end{equation}
where $\zeta_{n}$ are complex parameters and the  index $i$ denotes the sites.  Actually, we only need ten real parameters per molybdenum site and can for example set $\zeta_1=1$. The complex coefficients of six state would give 12 real parameters, but due to normalization and a variable global phase we are left with ten independent real parameters.

As there are four Mo ions in the unit cell of the fcc lattice we have all together 40 variational parameters in the site-factorized wave function $\prod_{i=1}^4|\psi_i\rangle$. Taking all 24 direction-dependent bonds of the unit cell as well as the spin-orbit coupling into account, we minimize the energy for these 40 variational parameters and compare it to the spin-orbital dimer-singlet solution. The resulting  phase diagram is shown in Fig.~4 of the main text.

\bibliographystyle{apsrev4-1}
\bibliography{Ba2YMoO6}

\clearpage

\end{document}